\documentstyle[epsfig]{elsart}
\begin{document}
\begin{frontmatter}
\title{Evolution and Ageing}

\author{S. Moss de Oliveira \thanksref{moss}}
\address{Laboratoire de Physique et M\'ecanique des Millieux 
H\'et\'erog\`enes, \'Ecole Sup\'erieur de Physique et de Chimmie 
Industrielles de la Ville de Paris, 10, rue Vauquelin, 75231 Paris Cedex 05, 
France}

\author{Domingos Alves} 
\address{Instituto de F\'{\i}sica de S\~ao Carlos, USP, Caixa Postal 369, 
13560-970 S\~ao Carlos, SP, Brazil}

\author{J.S. S\'a Martins}
\address{Colorado Center for Chaos and Complexity, CIRES, CB 216, 
University of Colorado, Boulder, Colorado, USA, 80309.}

\thanks[moss]{Corresponding author, presently and permanently at
{\it Instituto de F\'{\i}sica, Universidade Federal Fluminense,
 Av. Litor\^anea s/n, Boa Viagem, Niter\'oi 24210340, RJ, Brasil.
 E-mail address}: suzana@if.uff.br}

\begin{abstract}

The idea of this review is to connect the different models of evolution to 
those of biological ageing through Darwin's theory. We start with the Eigen 
model of quasispecies for microevolution, then introduce the Bak-Sneppen 
model for macroevolution and, finally, present the Penna model for biological 
ageing and some of its most important results. We also explore the concept of 
coevolution using this model.  

\end{abstract}


\end{frontmatter}

\section{Introduction}

Talking about biological ageing may be somewhat distressing, specially for 
those who are over forty, as are two of the authors. However, ageing is just 
one of the features of evolution, and to understand such an unavoidable 
mechanism it is necessary first to understand the paths of evolution. In 
fact, the concepts of evolution and ageing are connected by Darwin's theory 
of selection of the fittest, published in 1859. Insofar that it concerns the 
general 
evolution mechanism, this famous theory states that: {\it If genetically 
distinct individuals compete for limited resources, those more fitted to the 
environment will produce more offspring. Random mutations mix the genes, 
giving rise to new genetic combinations, and at every generation natural 
selection eliminates the less efficient ones, in order to continuously 
improve adaptation}.
However, since ageing is related to an age structure, the same theory 
concerning this subject is better enunciated as: {\it A mutation endangering 
the life of an organism before the reproductive age is much more dangerous to 
the species than a mutation affecting it only late in life, when it has 
already produced enough offspring to warrant the perpetuation of its 
lineage}. In both cases, the notion of fitness, one of the most debated 
quantities in population genetics, first introduced by Fisher \cite{Fisher} 
and Wright \cite{Wright}, seems to be closely related to the reproductive 
rate. Selection is also related to competition, which implies that the 
weaker are supposed to die. Thus, nature must offer a death mechanism, 
which turns to be the ageing process. 

Perhaps evolution can be fully explained by Darwin's theory. There is, 
however, a problem: where is the equation that would allow biologists, 
geneticists, mathematicians and even {\it exotic} physicists to understand    
how nature works and to publish more and more papers? Because such an 
equation doesn't 
exist, different models have appeared in order to explain the origins of 
life and its evolution. According to Luca Peliti \cite{Peliti}, these models 
can be divided into three groups. The first one concerns microevolution, that 
is, the evolution of individuals belonging to the same species or to closed 
ones. One example is the Eigen model for quasispecies \cite{Eigen}, described 
in section 2. In such models the interaction among individuals is generally 
introduced through some mechanism of global competition.  

The second group concerns coevolution, where two or more species 
interact strongly in such a way that the survival of one species depends on 
the survival of the other. The most common problem studied with these models 
is the prey-predator one, presented in a quantitative way by Lotka 
\cite{Lotka} and Volterra \cite{Volterra} many decades ago. Another example 
is the host-parasite interaction, which will appear later in this paper 
(section 4.4) as one possible explanation for the evolution of sex. 

Finally, the third group corresponds to models for macroevolution or 
large-scale evolution, that deal with all species alive at the same 
time but with no particular interacting mechanism between them. The concept 
of fitness for such models cannot be simply related to the reproduction rate 
of the individuals, since each species has its own reproductive strategy. It 
is necessary to consider instead a continuously evolving 
{\it fitness landscape}, that changes whenever a species mutates or 
disappears. In order to survive, the remaining species are also continuously 
evolving, trying always to be close to the peaks of the landscape. The 
Bak-Sneppen model \cite{Bak-Sneppen} is one of the simplest and most famous 
models of this third group. It considers the dynamics of large-scale 
evolution as a result of a {\it self-organized critical process}, responsible 
for the presence of scaling laws in macroevolutionary data.  

If it is easy to group the different models for evolution, the same is not 
true when the subject is ageing. In 1990 the Russians gerontologist Zhores 
Medvedev \cite{Russo} classified more than 300 theories constructed in  order 
to explain {\it the reduction of survival probability with advancing age}. 
Fortunately, these theories can be grouped in two types: the theories of 
``Why'' and the theories of ``How''. The first type tries to  understand 
ageing through a global perspective and to explain its disparities among the 
species. Theories of the  second type search for the specific mechanisms or 
immediate causes of ageing, each one with a given degree of validity 
depending on the species. 

The theories of ``Why'' are also divided in two groups. The first one 
attributes physiological unavoidable reasons for ageing, such as the 
harmful action of the oxygen radicals that are constantly produced in our 
bodies or the existence of a programmed cell death after a given limited 
number of cell divisions. The rate of metabolism of each species is probably 
the oldest example of this kind of theory. Since large mammals such as  
elephants move slowly and live much longer than small ones, which are very 
active and so have a high metabolic rate, for many years it was believed 
that this was the explanation for the disparities in the longevity of 
different species. However, it was discovered that there are branches of the 
same species that hibernate in the northern hemisphere and live as long as 
those branches that inhabit the southern hemisphere. There are also some 
kinds of birds that have an extremely long lifetime, despite the large amount 
of energy they need for flying. 

Observations like the ones above have gradually increased the importance of 
the second group of the ``Why'' theories, that is  
{\it The Evolutionary theories}, proposed around 1950 by Peter B. Medawar, 
George C. Williams and other famous biologists (for a review on the history 
of ageing theories and many of its interesting features see the special issue 
of La Recherche, 322, July/August 1999). These are theories based in Darwin's 
proposal of selection of the fittest \cite{Rose}. As mentioned before, an 
important ingredient for modeling using the evolutionary approach is to have 
an age structure with no reproduction during youth. One of the pioneers, and 
one of the simplest models using this strategy is the Partridge-Barton model 
\cite{Partridge-Barton}, which considers only two age intervals, one from t=0 
to t=1 for juveniles and another, from t=1 to t=2, for adults. 
Reproduction is possible only at ages 1 and 2 and is followed by death after 
age 2. However, when only deleterious inherited mutations are considered, 
this model leads to population meltdown, that is, the population dies out due 
to the accumulation of such mutations. The Penna model for biological ageing 
\cite{penna}, described in section 4, is also based on the mutation 
accumulation hypothesis and is now by far the most used one to reproduce and 
understand different aspects of real population dynamics. It deals with many 
age intervals and has successfully explained the catastrophic senescence 
of salmon; why women live longer than men; why does menopause exists, and 
other biological phenomena. Besides giving results that are in agreement with 
the empirical Gompertz law of an exponential increase of mortality with 
age, it doesn't lead to the population meltdown mentioned above.

This review is organized in the following way: the Eigen model is described 
in section 2, the Bak-Sneppen model in section 3 and the Penna model and its 
main results in section 4. In section 5 we present our conclusions.

\section{The Eigen Model for Microevolution}

The Eigen's quasispecies model \cite{Eigen} is one of the archetypical 
representation of the Darwin's postulates of evolution by natural selection. 
The model, originally developed to address the issue of explaining the origin 
of life on Earth, describes the dynamics of populations of replicating 
biological macromolecules under the influence of selection and mutation 
mechanisms. 
It predicts that if mutations occur frequently enough, the target of selection 
will no longer be a single individual, but rather an ensemble of genetically 
related individuals called a {\it quasispecies}. Moreover, when the mutation 
rate surpasses a critical value, known as the {\it error threshold}, 
the result is a complete loss of this polymorphic genetic structure. 
In this section we 
shall restrict ourselves to Eigen's first theoretical model of molecular 
evolution based on deterministic chemical kinetic theory (see \cite{Eigen89} 
for a complete review of the original formulation). Its conceptual elements 
and results may nevertheless be fairly well described, and perhaps even more 
appropriately, in the language of population genetics \cite{Alves96,Higgs94}, 
in terms of the frequencies of haploid multi-locus individuals (see 
\cite{Baake} for an excellent review on this subject).

\subsection{The Chemical Ansatz}

The original formulation of Eigen focus on a well-defined model system: the 
flux reactor. It comprises basically a reaction vessel, in which
biological macromolecules are continually built up out of energy-rich $T$ 
monomers (triphosphates), that are required for macromolecular synthesis, 
and decay, after a certain time, back to their energy-deficient $M$ 
monomers (monophosphates). Furthermore, it is assumed that this system can 
exchange energy and matter with its surroundings, by regulating both the 
supply of energy-rich and energy-deficient monomers, as well as the total 
population of macromolecules. Each chemical component of this reaction system 
consists of a reproducing macromolecule modeled as a single string of $L$ 
digits 
$I_{i}=\left( s_{1}^{i},s_{2}^{i},\ldots ,s_{L}^{i}\right)$, with the
variables $s_{\alpha }^{i}$ $(i=1,...,\kappa ^{L};\alpha =1,...,L)$ allowed
to take on $\kappa $ different values. Each of these values represents a
different type of monomer used to build the molecule (among biological
macromolecules $\kappa =4$ for nucleic acids G,A,C,T; $\kappa =2$ purines
and pirimidines; $\kappa =20$ for proteins). It is reasonable, therefore, 
to ignore the genotype-phenotype distinction; since the relevant Darwinian 
entities are replicating macromolecules, genotype and phenotype are aspects 
of one and the same object, its molecular sequence
(see \cite{Peliti} for a comprehensive overview). In this way, for one
particular string (or macromolecule, or genome) of type $i$, the various
events that could happen inside the flux reactor can be visualized as the
following single chemical reaction steps,

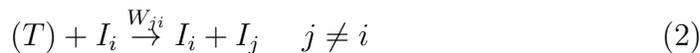
\begin{equation}
(T)+I_{i}\stackrel{W_{ii}}{\rightarrow }2I_{i}
\label{rea1}
\end{equation}

\begin{equation}
(T)+I_{i}\stackrel{W_{ji}}{\rightarrow }I_{i}+I_{j}~~~~j\neq i
\label{rea2}
\end{equation}

\begin{equation}
I_{i}\stackrel{D_{i}}{\rightarrow }(M)
\label{rea3}
\end{equation}

\begin{equation}
I_{i}\stackrel{\Phi _{0}}{\rightarrow }0.
\label{rea4}
\end{equation}

In these reactions, it is assumed that the available amount of $M$ and $T$
monomers is constant. The reaction (\ref{rea1}) denotes the self-replication, 
or error-free replication, of molecule $I_{i}$, and the reaction (\ref{rea2}) 
takes on the erroneous replication, or mutation, of $I_{i}$ that can lead to 
a new molecule $I_{j}$. The replication matrix $W$ takes into account the
primary structure of the macromolecules. This feature is what distinguishes
this model, a sequence space model, from models of population genetics \cite
{Baake,Alves97}. More specifically, its elements are given by 
\begin{equation}
W_{ii}=A_{i}\,q^{L}
\end{equation}
and 
\begin{equation}
W_{ij}=\frac{A_{j}}{\left( \kappa -1\right) ^{d\left( i,j\right) }}%
\,q^{L-d\left( i,j\right) }\left( 1-q\right) ^{d\left( i,j\right) }~~~~i\neq
j,
\end{equation}
where $A_{i}$ is the replication rate (or fitness) of molecules of type $i$, 
which tells us how fast new $I_{i}$ are synthesized in the next generation, 
independently of
whether the copies are correct or not. The parameter $d\left( i,j\right)$
is the Hamming distance between strings $i$ and $j$, i.e., the number of
monomers or positions at which these two sequences differ. Here, $q\in
\lbrack 0,1]$ is the fidelity parameter of each monomer, i.e., the
probability of inserting the correct monomer for any position.
Correspondingly $\mu =(1-q)$ is the error rate (or mutation rate) per
monomer, which is assumed to be the same for all monomers. Thus,
self-replication and mutation are represented by two autocatalitic reactions,
homogeneous and heterogeneous, respectively. The chemical decomposition of
the molecular specie $I_{i}$ is represented by reaction (\ref{rea3}), so
that $D_{i}$ is a general decay rate parameter. The reaction (\ref{rea4})
represents the efflux of the molecule $I_{i}$ by diffusion, where $\Phi_{0}$
is a global diffusion flux that is assumed to be the same for all molecules.

The relevant variables of this dynamical system are the concentrations of
each macromolecule $x_{i}=\left[ I_{i}\right] $. Then, inside the
reactor, the concentrations $x_{i}$ of molecules of type $i=1,2,\ldots
,\kappa ^{L}$ evolve in time according to the following differential
equations 
\begin{equation}
\frac{dx_{i}}{dt}=\sum_{j}W_{ij}x_{j}-\left[ D_{i}+\Phi _{0}\right] x_{i}.
\label{ODE}
\end{equation}
The quasispecies model can thus be described as a population dynamics model 
given by the full set of phenomenological differential equations. 

The equations (\ref{ODE}), however, do not yet lead to \textit{competitive
selection} in our model system. Moreover, in order to induce selection
pressure in the system described by (\ref{ODE}) it is necessary to impose
some type of restriction to it. To appreciate the effect of the competition
in the model system (that may or may not lead to selection), let us imagine the
ideal situation where the molecular species reproduce themselves without
error, so that $\mu =0$ and $W_{ij}=0,$ $\left( i\neq j\right) $. In
this case, equations (\ref{ODE}) become 
\begin{equation}
\frac{dx_{i}}{dt}=\left[ W_{ii}-D_{i}-\Phi _{0}\right] x_{i}.
\label{ODE1}
\end{equation}
It is trivial to see that the solutions for the molecular concentrations are
exponentials if $\Phi _{0}$ is constant. Thus, if there are $k$ distinct 
species of molecules inside the reactor, the concentration of all species 
with replication rate $W_{kk}>$ $D_{k}+\Phi _{0}$ grows exponentially  as 
time goes by, while all those with $W_{kk}<D_{k}+\Phi _{0}$ die out. In this 
scenario, there is no competition and thus no selection. This behaviour of 
\textit{segregation} in the model is show schematically in Figure 
\ref{fidom1}(a) for the case of five types of binary molecules 
($\kappa =2$, $s_{k}={0,1}$), all with the same frequencies at $t=0$.

In this case of replication without error, we can now subject the molecular
population inside the reactor to some form of global constraint. This
can be made by keeping the total population of molecular species constant, 
$\sum\nolimits_{i}x_{i}=N$. For this purpose the global dilution flux 
$\Phi_{0}$ will be adjusted in time so as to keep pace with the increase in 
total molecular concentration, i.e., it will be determined by the condition 
$\sum\nolimits_{i}dx_{i}/dt=0$. Therefore, the global flux of molecules
inside the reactor must satisfy the condition 
\begin{equation}
\Phi _{0}=\frac{\sum\limits_{i}(W_{ii}-D_{i})x_{i}}{N}.
\label{prod}
\end{equation}
Now, with this equation for $\Phi _{0}$, the set of differential equations 
(\ref{ODE}) and (\ref{ODE1}) is non-linear. In Figure \ref{fidom1}(b) the 
temporal behavior of the same molecular populations of example of Figure 
\ref{fidom1}(a) is 
schematically shown when we constrain the system to have a constant
population. In fact, as $\Phi _{0}$ increases due to the constant
population condition, a higher number of molecules is segregated, and only 
the one with the highest productivity will survive. In the stationary state 
of this competition, the winner is called the \textit{master sequence }
($I_{m}$) and the stationary state of this system is termed 
\textit{selection equilibrium}. The molecular selection process is, 
therefore, an environmental effect, through a constraint imposed on molecular 
population, and always leads to an unambiguous selection decision.

\subsection{Quasispecies and error threshold}

Let us discuss briefly the more general case, described by equation 
(\ref{ODE}). If we allow for erroneous replication ($\mu >0$), considerations 
similar to the ones above lead to the following condition on $\Phi _{0}$ 
\begin{equation}
\Phi _{0}=\frac{\sum\limits_{i}\sum\limits_{j}W_{ij}x_{j}-\sum%
\limits_{i}D_{i}x_{i}}{N}.
\label{prod1}
\end{equation}
The selection equilibrium solution of (\ref{ODE}) can be found in terms of
the eigenvalues and eigenvectors of the replication matrix $W$. If we define
the vector $\mathbf{y}=(y_{1},y_{2},...y_{L})$, where the components
represent the relative concentrations (or frequencies) of each molecular
species in the population, $y_{i}=x_{i}/\sum\nolimits_{j}x_{j}$, we can write 
\begin{equation}
W^{\prime} \mathbf{y}=\lambda \mathbf{y},
\end{equation}
where the diagonal elements of $W$ were modified to 
$W_{ii}^{\prime}=W_{ii}-D_{i}$ (see the appendices of \cite{Eigen89} for 
details on how to 
transform the non-linear system described by (\ref{ODE}) and (\ref{prod1}) in
a linear one). Then, we define a quasispecies precisely as the dominant
eigenvector $\mathbf{y}_{\max }$ associated to the largest eigenvalue 
$\lambda _{\max }$ of the replication matrix $W^{\prime }$. This eigenvector
describes the exact population structure of the population: each mutant 
$I_{i}$ is present in the quasispecies with frequency $y_{i}$ 
($\sum\nolimits_{i}y_{i}=1$). Note that the largest eigenvalue is exactly 
the average replication rate of the quasispecies, $\lambda _{\max
}=\sum\nolimits_{i}A_{i}y_{i}$. In this scenario, the frequency of a given
mutant within the quasispecies does not depend on its replicative value
alone, but also on the probability with which it is produced due to erroneous
replication of others molecular species. Moreover, depending on the mutation
rate $\mu =(1-q)$, the master sequence and some (or all) mutants coexist as a
quasispecies. Then, one of the main outcomes of this model is that, in 
equilibrium, selection does not, in general, lead to a homogeneous population 
formed by a single kind of molecular individual: a particular sequence is no 
longer \textit{the outcome} of selection. A \textit{set} of genetically 
distinct sequences, forming a mutant distribution centered around the master 
sequence, is produced instead. 

The mutation rate in this model is the parameter that controls the width of
this distribution, that is, how much the quasispecies spreads over the
space of sequences. Moreover, one immediate consequence of the existence of
a maximum eigenvalue is the appearance of a threshold relation. Then, for the 
binary version of this model, as the error rate $\mu $ increases, two
distinct regimes are observed in the population composition: the \textit{%
quasispecies} regime, characterized by the master string and its close
neighbours, and the \textit{uniform} regime, where the $2^{L}$ possible 
sequences appear in the same proportion. The transition between these
regimes takes place at the \textit{error threshold} $\mu _{t}$, whose value
depends on the parameters $L$ and $A_{i}$ \cite{Eigen,Eigen89}. This becomes 
a genuine phase transition from an adaptive to a disordered neutral 
phase in the limit $L\rightarrow \infty $ \cite{Peliti}.

Hence, to proceed further we must specify the replication rate of each
sequence, or, in a more concise form, the \textit{fitness} or 
\textit{replication landscape} (this term was originally coined by S. Wright 
\cite{Wright2}), a line drawn by all $A_{i}$ points related to the sequence 
space.
In particular, the quasispecies concept and the error threshold phenomenon
are illustrated more neatly by the so-called \textit{single-sharp-peak}
landscape. This simple, and probably the most studied of the landscapes, can 
be constructed by ascribing the replication rate $a>1$ to the master sequence 
$\left(1,1,\ldots ,1\right)$. The remaining sequences, differing by as little 
as a single mutated monomer, are at a disadvantage $a-1$ as expressed by their 
replication rate $a'=1<a$. In Figure \ref{fidom2} we present the steady-state
molecular frequencies in the case where $L=30$ and $a=10$, as a function of
their Hamming distances $d=0,\ldots ,L$ from the master sequence. That is, the 
$2^{L}$ possible sequences were grouped into $L+1$ different classes of
sequences, characterized solely by the number of mutant monomers 
($s_{\alpha}={0}$) they have, regardless of the particular positions they 
occupy inside the sequences. Then, if we take as a reference the case 
$\mu =0$, where only the master sequence survives in the selection 
equilibrium, the parts (a) to (c) of the figure show three situations in the 
quasispecies (or adaptive) regime, $\mu <\mu _{t}$: in (a) the master sequence 
is the more frequent, in (b) the master sequence is no longer the more 
frequent, and in (c), near the error threshold, the number of master copies 
becomes strongly reduced, but the quasispecies is evolutionarily most 
versatile because it produces a wide variety of mutants without destabilizing 
the master sequence. Part (d) illustrates the uniform (or stochastic) 
regime, $\mu >\mu _{t}$, where the $2^{L}$ possible genomes appear in the 
same proportion or the $L+1$ classes are distributed by the binomial 
distribution $y_{d}=\frac{1}{2^{L}}{L \choose d}$.

Therefore, in this simple example of fitness landscape, without errors ($\mu
=0$) or with too much errors ($\mu >\mu _{t}$), there is no evolution. In
the first case there are no mutants; in the second case there is no
adaptation. Moreover, it is possible to show that the existence of a
threshold relation, as pointed out above, correlates the mutation rate with a
maximum length of the sequences that can be reproducibly maintained by
selection (see \cite{Smith} for a simplified case, but with a very intuitive
appeal). For the single-sharp-peak landscape: 
$$L<\frac{lna}{1-q}$$
That is, once the fidelity degree $q$ of the copying mechanism is fixed, the
molecule size cannot exceed a given value $L_{max}$. On the other hand, it is
impossible to increase the fidelity of the copying mechanism without
changing $L$; this is the so-called Eigen's paradox. It poses a serious 
difficulty in envisioning life as an emergent property of systems of 
competing self-replicating macromolecules.

Some final comments regarding the formulation described above are in order.
It is valid only in the limit where the total number of molecules $N$ goes
to infinity. Of course, all real populations are \textit{finite}, and they
will not behave in the deterministic way expected for an infinite population 
\cite{Alves98}. As pointed out by Higgs \cite{Higgs96}, the steady-state of
a finite population is a dynamic one in which the population can continue to
evolve, and therefore it is not equivalent to an infinite population model. 
Finally, it should be added that the error threshold itself is not a general
phenomenon, and can be absent even for landscapes as simple as the Fujiyama 
one \cite{Peliti} (for a lucid discussion on the limits and further 
developments of this model see Ref.\cite{Baake} and references therein). For 
all other landscapes, such as the non-stationary one described in the next 
section, more elaborate approaches are required to clarify the real meaning 
of the error threshold. As a side remark, and to establish another 
connection with what follows, the evolution of a finite population in a 
random fitness landscape has been recently shown to exhibit a punctuated 
scenario \cite{Zhang}. 

\section{The Bak-Sneppen Model for Macroevolution} 

In his excellent book ``How Nature Works'' \cite{Bak-livro} Per Bak argues 
that: {\it Complex behaviour in Nature reflects the tendency of large systems 
with many components to evolve into a critical state where minor disturbances 
may lead to events, called avalanches, of all sizes. This delicate state 
evolves without interference of any outside agent and so is a critical 
self-organized state, that appears as a consequence of the dynamical 
interactions among individual elements of the system}. Self-organized 
criticality is now a widespread concept and many different systems  are known 
to evolve according to this dynamics. Particularly, the sandpile model 
\cite{Bak-livro,sandpile} is the best one to explain what the famous 
avalanches are and how to measure them. Due to lack of space, we are forced 
to go 
directly to the Bak-Sneppen model and to the avalanches that appear in the 
large-scale evolutionary process that can explain, for example, the mass 
explosion of the Cambrian period and the mass extinction of the Cretaceous, 
when the dinosaurs disappeared.

In this model there are $I$ species, each one occupying one site of an 
unidimensional lattice (a ring). Each species has a random fitness 
$0 \le f_i \le 1$. The simulation evolves according to the following rule: 
\begin{itemize}
\item{search for the smallest $f_i$ corresponding to species $i$;}
\item{change $f_i$, $f_{i-1}$ and $f_{i+1}$ for 3 other values randomly 
chosen (mutation or extinction);} 
\item{return.}
\end{itemize}

At the start of the simulation the fitness on average grows, although there 
are fluctuations up and down. However, after a transient period, the system 
reaches a stationary critical state where the fitness does not grow any 
further 
on average: {\it all species have fitness above some threshold} very close to 
$2/3$ (state of ``stasis''). Consider a point in time when all species are 
over the threshold; at the next step the least fit species (right at the 
threshold) will be selected, eventually starting an avalanche  or 
``punctuation'' of mutation events. (In fact, whenever the fitness of a given 
species changes, it is possible to think that the species has undergone a 
mutation or that it has become extinct.) After a while, the avalanche stops, 
when again all species have fitness above the threshold. The avalanche size 
corresponds to the number of steps needed to recover the state of stasis (or 
equivalently, to the number of active species between two consecutive states 
of stasis). When this process is repeated for large systems (large number of 
species), one obtains the number $N(S)$ of avalanches of size $S$ given by the 
power law: 
$$N(S) \propto S^{- \tau} .$$
This distribution means that there is no characteristic size for the 
avalanches, as would happen if instead of a power law it had an exponential 
behaviour. The larger the system, the larger the possible maximum size of 
an avalanche is. Small avalanches, in which a few number of species become 
active, are much more frequent than large ones; however, the probability that 
a system-sized avalanche occurs, activating all species, is not zero. Such a 
dynamical behaviour can explain the extinction of the dinosaurs without using 
any external agent, such as meteorites colliding with Earth (but of course 
does not exclude such a possibility). 

The important point here is the fact that a given species, though highly fit, 
can be chosen to mutate because one of its neighbours has a low level of 
fitness. In this sense the species live in a continuously 
evolving fitness landscape, differently from the quasispecies of the Eigen 
model in which the fitness landscape is fixed: once a replication rate is 
attributed to a given macromolecule, it never changes.    

Finally, in the next section we introduce the Penna model for biological 
ageing. It is also a model for microevolution, where individual interactions 
are introduced through a competition for food and space as in the Eigen 
model, but one that presents an age-structure, without reproduction at early 
ages.

\section{The bit-string Penna model}

The most successful computational model for age-structured populations is by 
far the Penna model \cite{penna}. One of the reasons for its success relies 
on a particularly well-suited computational representation of a genome by 
means of a sequence of bits, the bit-string. When grouped into computer words 
these strings can be efficiently operated on by very fast logical and bit-wise 
CPU instructions. Conceptually, the model is extremely simple, but the results 
it shows are far from trivial. It has also proved to be flexible enough to 
be of value in a number of different problems in population dynamics, and the 
recent literature is eloquent proof of this statement.

The biological support for the Penna model comes from the mutation 
accumulation theory of senescence \cite{medawar,gill}. In essence, this theory 
relates the evolution of senescence to the action of age-specific deleterious 
mutant alleles and the maintenance of some of these genes by the combined 
effects of mutation and selection pressures. In response to this conflict, 
deleterious mutant alleles with late ages of action would have higher 
equilibrium frequencies of affected individuals than genes with similar 
effects on survival early in reproductive life \cite{cworth}. A sufficiently 
large number of loci capable of mutating to deleterious alleles with 
age-specific effects would therefore generate a net decline in survival 
with advancing adult age. 

\subsection{Haploid asexual population}

We will begin the description of the Penna model with its simplest 
implementation, the one that addresses haploid asexual populations. For a 
much more detailed explanation, together with a sample computer code, the 
reader is directed to Ref. \cite{teubner}. The basic 
structure of the model is the age-structured genome pool. Each genome from 
this pool is associated to an individual, and its pattern of alleles 
summarizes the genetic heritage acquired by this particular individual. The 
only genes represented are those with age-specific effects. Each gene appears  
in a particular position (locus) of the bit-string, and this position is 
associated with the age from which it becomes effective. A gene can appear in 
two alleles, for the $2$ possible values of the bit at one locus: a bit set 
to $1$ represents the deleterious allele, and it is set to $0$ for the 
non-deleterious variety.

This basic and simple structure allows the computation of the number $A$ of 
active deleterious mutations at any point of an individual life time: one has 
only to add up the bits of the bit-string from the first locus to the one 
corresponding to the actual age of that individual. The size $S$ of the 
bit-strings is the first parameter of the model, and determines the maximum 
theoretical age of the individuals. In this context, age is not to be 
understood as measured in human years, but rather in a species-specific time 
unit.

Selection pressure is modeled by the introduction of a threshold $T$ for the 
number of deleterious mutations that can be simultaneously active in a living 
individual's genome. Usually, the probability of death because of genetic 
causes is assumed to be given by a step function $\Theta(T - A)$, with 
$\Theta(x)=1$ if $x \leq 0$, which introduces a high degree of non linearity 
in the model. Smoother functional forms have also been used \cite{fermi}, but 
the fundamental results of the Penna model do not appear to be sensitive to 
this choice.

Death for non-genetic causes, representing the outcome of intra-species 
competition for the limited resources of the environment, is modeled by a 
density- and time-dependent quantity, the Verhulst factor. This is a 
mean-field death probability, given by $V(t) = N(t)/Popmax$, where $N(t)$ is 
the total population at the beginning of time step $t$ and $Popmax$ is another 
parameter of the model that quantifies the above mentioned environmental 
constraints on the size of the population. The introduction of the Verhulst 
factor, or some equivalent form of limiting factor for the total population, 
is a necessity in simulational models to avoid population overflow, although 
its biological motivation and implementation strategy have recently raised 
some interesting questions \cite{cebrat}.

In an asexual population all the individuals are female. After reaching some 
minimum age $Minage$, and until she is $Maxage$ time periods old, every 
female in the population gives birth to $B$ offspring at every time step. At 
this moment, the genetic heritage of the mother is copied to her offspring, 
and mutations can occur. This process is simulated by generating, for each of 
the $B$ offspring, a genome cloned from that of the mother. On this 
bit-string, $M$ mutations are introduced in randomly chosen loci. It is usual 
to consider only deleterious mutations, since they are the overwhelming 
majority in nature. A clever coding trick is in order at this point: for the 
$\Theta$ function implementation of the rule for genetic deaths, one might as 
well compute at this moment the programmed age of death, by adding up the 
value of the bits from locus $0$ until this sum reaches the threshold $T$; 
the last locus to be computed in this sum corresponds to the age at which this 
individual will die, if not sooner because of the Verhulst dagger. This trick 
avoids having to compute, at each time step, the number of active deleterious 
mutations for each individual, a time-consuming and redundant operation.

With the elements above, a typical simulation of the Penna model undergoes 
the following steps, described in a simple auto-explanatory meta-language:
\begin{itemize}
\item An initial population of $Inipop$ individuals is generated. The usual 
choices are either a mutation-free population, in which all the bits of all 
the bit-strings are $0$, or one in which each individual has a random number 
of mutations, between $0$ and $S$, in randomly chosen loci.
\item FOR each time step:
\item COMPUTE the Verhulst factor $V$;
\item FOR each individual:

\indent INCREMENT age by $1$;

\indent IF age is smaller than the programmed age of death AND a randomly 
tossed number in the interval $(0,1)$ is larger than $V$ THEN reproduce, 
giving birth to $B$ mutated clones;

\indent ELSE she dies: her genome is erased from the genome pool.
\end{itemize}
These dynamic rules are executed for $Nsteps$ time steps. In the last 
$Asteps$ averages are taken over the population, and constitute the outcome 
of the simulation. The quantities usually computed are:
\begin{itemize}
\item The age distribution of the population, i.e., a histogram of the number 
of individuals at each age, with some normalization.
\item The survival probability as a function of age, defined as the ratio 
$N(a+1,t+1)/N(a,t)$ between the population with age $a+1$ at time step $t+1$ 
and the population that at the previous time step had age $a$.
\item The mortality rate, defined as the logarithmic measure of the 
population decay, normalized so that its value at age 1 is zero and where 
only deaths due to genetic causes are considered.
\item The genetic state of the population, which can be measured for instance 
by the fraction of defective genes in the population at each locus.
\end{itemize}

\subsection{Early results}

Among the many results obtained with the Penna model already published, we 
chose two of the most spectacular to comment on. The first one shows the 
agreement of the simulation results for the mortality rate with the 
empirically derived Gompertz law \cite{gompertz}. This law was proposed in the 
late $19$th century to account for the observed mortality rate of the German 
population, and has since then been verified by a number of observations of 
both human and mayfly populations. It states that the increase of mortality 
with age is exponential. The Penna model was the first computational model 
for ageing that could reproduce this result. In Figure 
\ref{figomp}
we show data derived from a simulation - the particular parameters used can 
be found in the caption. 

The second result deals with the catastrophic senescence observed in 
semelparous species. These are species that reproduce only once in a lifetime, 
such as the pacific salmon. Ageing in these species is called catastrophic 
because the females die soon after giving birth for the first and only time. 
To simulate a semelparous population with the Penna model one only needs to 
set $Minage = Maxage$, as opposed to having $Minage < Maxage$, which is the 
normal (iteroparous) case. These simulations could show that the catastrophic 
senescence effect is due to the lack of selection value of age-specific genes 
that become effective after the reproduction age \cite{salmon}. With the 
absence of selection, the mutation pressure turns on deleterious alleles for 
all these genes. In the computational representation provided by the Penna 
model, all loci associated with ages greater than the reproduction age become 
set to $1$ and the individual dies because of the accumulation of deleterious 
mutations in the first time step following its progeny. The consequences 
of this pattern of fixation of deleterious alleles can be clearly seen in the 
plot for the survival rate shown in Figure 
\ref{fisalmon}
which compares semelparous and iteroparous populations with equal parameters. 
The survival rate for the semelparous population has an abrupt decay to zero 
at age $= Minage+1$, in sharp contrast with the iteroparous one. Setting 
$Minage = Maxage$ also greatly simplifies the analytical formulation of the 
Penna model and allowed a formal derivation of the catastrophic senescence 
effect \cite{analit}.

The analysis of Figure 
\ref{fisalmon}
-b suggests an interesting puzzle. The maximum age of an individual in this 
population is its last age of reproduction: {\it there is no post-reproductive 
life}!. The conflict between selection and mutations has a very simple 
outcome, and the individuals die as soon as they loose their function of 
perpetuating the species. This is not what is seen in nature, however. In 
human and some other mammal populations, females live after having ended 
their reproductive period (after menopause). To address this question the 
Penna model needs an extension to sexual diploid populations.

\subsection{Diploid populations}

These are species where the genetic information is carried by two homologous 
strains. The effectiveness of a deleterious allele at one locus now depends 
on a combination of information carried by the two strings. The concept of 
dominance appears in connection with this issue. If a dominant allele appears 
in one locus of any of the two strings, it is effective, irrespective of the 
allele that is present in the homologous locus of the other. For a 
non-dominant allele to be effective, on the other hand, it must be present in 
both homologous loci. 

The appearance of diploid organisms in the life story of our planet marks also 
the onset of recombination and sex in the process of reproduction 
\cite{maynard}, although there are quite a few examples in nature of asexual 
reproducing diploid species where recombination is also present. This last 
strategy of reproduction, asexual but with recombination, is called meiotic 
parthenogenesis. Reproduction in diploid organisms involves the generation of 
a haploid cell, a process called meiosis, containing a subset of the genetic 
material of the parent that is to be transmitted to the offspring. To generate 
this haploid cell, alleles of the two homologous strains are reshuffled and 
recombined to form a single strain in a process called recombination. 
Diploidity and sex raise a whole new set of questions in population dynamics, 
some of which have been already studied in the framework of the Penna model 
and will be discussed in the sequel.

The extension of the Penna model to deal with diploid populations is rather 
straightforward \cite{diploid,diploid1}. Now, the genome of each individual is 
composed of two bit-strings to be read in parallel: two alleles, one from each 
string, have to be taken into account to decide if the character of each 
age-dependent gene is deleterious or not. If this is an homozygote locus, i.e. 
if both alleles have the same character, then the gene has also this 
character. If the locus is heterozygote, the decision has to respect the 
dominance rule. To implement this rule, an extra $S$ bits long bit-string is 
generated at the beginning of a simulation indicating, for each locus, which 
is the dominant character. In $D$ of the $S$ loci, randomly chosen from a 
uniform distribution, the deleterious character is dominant. For these loci, 
a $1$ bit set in any of the two strains suffices to define the character of 
this gene as deleterious. For the $S-D$ remaining loci, the deleterious allele 
is recessive, and the homologous bits of the two strings have to be both set 
to $1$ for any of these loci to have a deleterious effect.

Meiosis in the Penna model can be easily tailored to mimic nature. As a side 
example, apomyctic parthenogenesis, a diploid asexual mode of reproduction 
without recombination, would be represented by a simple random choice of one 
of the two genetic strains. For recombination, the usual is to select a random 
position out of the $S$ loci and cut the two strings at this position. Two new 
strings are generated by crossing the resulting four pieces: the left side 
coming from one of the strings is attached to the right side coming from the 
other. Of the two new strings, one is randomly chosen, and constitutes the 
genetic material to be inherited by the newborn. For meiotic parthenogenesis, 
this single strain is cloned: before mutations, the new genome is totally 
homozygote. For sexual species, a male individual is randomly selected and his 
genetic material also undergoes meiosis and recombination. The two strains, 
one coming from the female and another from the male, form the genome of the 
newborn. Its gender is now randomly chosen, with equal probabilities. On each 
of these two strings, $M$ (deleterious) mutations are added at randomly 
chosen loci. In Figure \ref{firegi} we illustrate the procedures for 
reproduction of haploid asexual, diploid sexual and diploid meiotic 
parthenogenetic populations with simple examples.

Now we are in a position that allows us to address the puzzle with which we 
ended the last subsection, namely, why do women live as long as men instead 
of dying immediately after their reproductive period? And we show in Figure 
\ref{fisurvsex}
the resulting survival rates when sexual reproduction is added to the model. 
Semelparous populations still suffer catastrophic senescence, even when the 
males do not loose their reproductive capacity at any age \cite{kathia}. But 
for iteroparous species, females now have a post-reproductive life and there 
is no difference between male and female survival rates. The 
presence of males in the population bring some benefits to the females, after 
all! But this observation does not ends our quest, for in the same Figure 
we also see that populations for which females do not undergo menopause, and 
are able to breed for their entire life, have a larger life span. Why would 
then nature ``invent'' menopause?

The answer to this question is now rather elaborate, and has been suggested 
already in the biological literature. It comes as a result of three new 
components, and the ability to acomodate and manipulate them is a 
demonstration of the flexibility of the model. These new components are the 
need for parental care of the newborn during a certain period of time, the 
reproductive risk that increases the death probability for a female at the 
moment of delivery, and the transformation of one of the parameters of the 
model, the maximum age of reproduction $Maxage$, in an individual and 
genetically acquired characteristic, now subject to mutations. The first new 
component is implemented in the model by requiring an infant to have a living 
mother in order to survive during its first $Apc$ periods of life. For the 
reproductive risk, a dependence on the number of active deleterious mutations 
$A$ was introduced. Thus, a female can die, with a probability essentially 
given by $A/T$, at the moment of delivery. If the newborn is a female, she 
inherits the same $Maxage$ of her mother with some probability $P$, or 
mutates to have it increased (decreased) by $1$ with probability $(1-P)/2$. 
The Darwinian dynamics of the model is sufficient to produce a 
self-organization of the distribution of the menopause age of the female 
population, showing that inhibition of reproduction after a certain age is 
actually beneficial for the species as a whole \cite{meno}. In Figure 
\ref{fimeno}
the distribution of the age of menopause onset throughout the population is
shown in a comparison between a population where parental care is not needed
and one where an infant aged below a minimum would die if the mother was not
any more present.

\subsection{The evolution of sex}

The maintenance of sexual reproduction among the great majority of species in
nature, in spite of the inherent cost of having to produce males to assure
reproduction, is still one of the great puzzles of biology. In fact, a simple
reasoning shows that the need of two parents to generate even a single
offspring should give asexual varieties a two-fold advantage over sexual ones
\cite{maynard}. The advantage of sex relies on its ability to create greater 
genetic diversity, since the pool of alleles from which the newborn genome 
is extracted is different for each mating pair. Asexual reproduction, on the 
other hand, generates new genomes from a more limited pool, since there is 
only one parent involved. It is not clear though in what circumstances this 
greater diversity provided by sex would give it the upper hand against, say, 
meiotic parthenogenesis. One could argue that the species would benefit from 
the cloning of well-fitted genome, only possible in asexual reproduction, and 
that sex would make these genomes short-lived for exactly the same reasons 
it can create diversity. Sex could possibly be more efficient in getting rid 
of bad mutations, by bundling them together through mating and expelling them 
from the genetic pool through selection. These are as yet open questions, and 
the Penna model has been used to address them, even when the age structure of 
the population is not the issue. This is possible because of its particularly 
simple implementation of a selection mechanism, which is a general requirement 
for any model to be useful in this context.

For the evaluation of diversity in the Penna model one measures the number of 
different alleles, or bits, for each pair of genomes in the population. The 
resulting distribution of this so-called Hamming distance has a Gaussian 
character for any reproduction strategy. The comparison between sex and 
meiotic parthenogenesis show a similar width for the distributions, but a 
higher value for the distance where it peaks in the sexual case \cite{why}. A 
particularly interesting reflex of this greater diversity can be seen when a 
genetic catastrophe is provoked at a chosen time step by instantaneously 
setting an extra deleterious mutation at a chosen locus in all the 
individuals. An equivalent, and perhaps more easily translated into 
biological terms, procedure would be to decrease the threshold of deleterious 
mutations $T$ by one, representing a sudden depletion of vital environmental 
life protection resources. Thanks to their greater diversity, sexual 
populations are always resistant to this catastrophe, whereas simple asexual 
reproduction leads invariably to extinction, as shown in Figure \ref{ficat}. 
For meiotic parthenogenesis, the outcome is not so clearly cut, but 
extinction is a possibility \cite{why}.

A number of theories have been put forth to try to explain the evolution and 
maintenance of sexual reproduction. In the center of this debate is the 
so-called ``Red Queen'' hypothesis, that relies heavily on the ideas of 
diversity discussed above. In essence, it holds the action of genetically 
matching parasites as responsible for creating a rapidly changing environment. 
In this unstable ecology, only varieties that can mutate their genomic pool 
at least as fast as the adaptation of the parasites proceed can survive. The 
theory derives its name from this endless race, quoting from the Red Queen of 
Lewis Carol's Alice in Wonderland: ``It takes all the running you can do, to 
keep in the same place.'' In fact, recent observations of competing varieties 
of a freshwater snail, {\it Potamopyrgus antipodarum}, have shown that there 
is a strong correlation between the prevalence of one reproduction regime and 
the concentration in its habitat of the trematode {\it Microphallus}, a 
parasite that renders the snail sterile by eating its gonads 
\cite{lively,lively1,lively2}. 
Namely, the asexual variety is predominant where the parasite appears in small 
concentrations, whereas higher concentrations of the trematode forces the 
species to prefer a sexual regime.

This correlation could be shown to exist in simulations of a conveniently 
modified Penna model. The parasites are represented by a dynamically changing 
memory bank of genomes of some fixed number of entries. Each entry is modified 
if it comes into contact with the same genome twice in a row; in this case, 
it memorizes this pattern and stores it in the memory bank. At each time step, 
before the reproduction cycle, each female of the population is probed by a 
fixed number $E$ of randomly chosen entries of the parasite bank. If one of 
these entries is a perfect match for the female's genome, she is rended 
sterile and can no longer reproduce. The number of parasite exposures $E$ is 
an indirect measure of the parasite concentration in the habitat. For the 
host population, the reproductive regime of the females is no longer a fixed 
character, but can mutate with some small probability. The simulations begin 
in the absence of the parasite infestation, and the initial population is set 
to have a sexual reproductive regime. As soon as the meiotic population 
appears, due to mutations in the reproductive regime, it overrides the sexual 
variety, in a demonstration of the two-fold disadvantage of sex above 
mentioned, and sex barely subsists due to infrequent back-mutations from the 
asexual variety. At some time step, the parasite infestation is turned on. 
The resulting predominant variety is going to depend solely on the intensity 
of this infestation, as measured by the exposure parameter $E$. For small 
values of $E$, the asexual variety has the upper hand. As $E$ is increased, a 
first-order transition is seen to a configuration dominated by the sexual 
population \cite{redq}. Figure \ref{firedq} shows the fraction of females in 
the population that reproduces sexually, as a function of the exposure 
parameter $E$. The sudden jump in this fraction signals the order of the 
transition.

\section{Other applications of the Penna model}

It is not difficult to find in the animal kingdom species that live and work 
in sexual pairs, but sometimes have an extra-pair relation, like the 
Scandinavian great reed warbler, chimpanzees, etc... (of course some 
disgusting men also belong to this category). As already shown by Martins and 
Penna \cite{Martins}, such a behaviour increases the genetic diversity and 
may lead to better fitted offspring depending on how females select the males 
for an extra-pair relation. However, as in the case of the parasites presented 
before, nature not always chooses the strategy leading to the highest 
reproduction rate. Another example is the California mouse \cite{mouse}, one 
of the rare monogamous species that have been found. These mouses live in a 
extremely cold place, and in order to survive the pups must be continuously 
heated by the body of one of the parents. When the male abandons the nest, 
the female very often kills the babies. Starting with a population with half 
of the males faithful and the other half non-faithful, Sousa and Moss de 
Oliveira \cite{fidelity} have shown that depending on the death probability 
of the abandoned offspring, the population may self-organize in a situation 
where all males are faithful despite of reproducing lesser. In their 
simulations, monogamy is paternally transmitted and exclusively related to 
parental care. In fact it is already known that there some genes responsible 
for maternal care that are paternally transmitted \cite{Mest1,Mest2}.  

Another interesting result obtained with the Penna model concerns the higher 
mortality of males when compared to the females one. The mortality curve as a 
function of age for females is lower since birth until advancing ages 
(around 90 years), when both males and females mortalities become equal. 
Stauffer et al \cite{somatic} introduced somatic mutations (that are not 
transmitted to the offspring) into the model, atributting to males a higher 
somatic mutation rate than to females. With this strategy they were able to 
reproduce the observed behaviour of the mortalities, with the females one 
lower than that of males until advancing ages, when the genetic mutations 
dominate and the two curves collapse. Penna and Wolf \cite{Wolf} obtained the 
same result atributting to the females a higher value of the limit number of 
genetic diseases $T$ than to the males. However, the best strategy was 
proposed by Cebrat \cite{Cebrat}, and confirmed by Schneider \cite{Schneider} 
et al., that modified the model in order to distinguish the double X 
chromosomes of the females from the single one of males. Considering the 
mutations in the single X male chromosome as dominant mutations, even more 
realistic results were obtained. 

Finally, we must say that it has also been possible to predict some unknown 
effects using this model. For instance, it has been shown that in small 
populations, if a given percentage of males is periodically substituted by 
the same percentage of males coming from a large population, the extinction 
of small populations due to inbreeding may be avoided \cite{inbreeding}. Also 
a nice strategy for fishing some species in which fertility varies with size 
in order to warrant a larger stock without loosing money, has been recently 
proposed by Racco and Penna \cite{Racco}.

\section{Conclusions}

We have shown that different evolutionary models found in the literature are 
connected to the biological ageing ones through the Darwin's theory of 
selection of the fittest. The models we have presented here are the most 
simple ones, but their results are far from trivial. The Eigen model for 
microevolution makes clear the difference between segregation and selection, 
as well as the connection between mutations and diversity. The Bak-Sneppen 
model exemplifies the importance of a continuously evolving fitness landscape 
to simulate large-scale or macroevolution.

Finally we have presented the Penna model for biological ageing and some of 
its most important results. Ageing is an unavoidable process (except for 
some {\it rare individuals} like D. Stauffer, who is eternally young) and has 
been extensively studied by many different scientists, since a very long 
time. Although the evolutionary theories for senescence have appeared around 
1950, Monte Carlo Simulations on this subject started only after the 
publication of the Partridge-Barton analytical mathematical model in 1993. 
The Penna model is now the most widespread Monte Carlo technique to simulate 
and study the different aspects of population dynamics, including ageing. In 
this review we have focused attention on results concerning the differences 
between reproductive regimes and the advantages of sexual reproduction 
(although our arguments have not proved good enough to convince the rare 
individual just mentioned above that males are all alike but still useful).

\section*{Acknowledgments}
S.M.O. and J.S.S.M. thank P.M.C. de Oliveira, D. Stauffer, T.J.P. Penna, 
A.T. Bernardes, A.O. Sousa and S. Cebrat for many discussions and 
collaborations in previous papers. 
S.M.O.'s work is partially supported by the Brazilian Agencies CNPq, Faperj 
and Capes. 
D.A.'s work is supported by FAPESP. 
J.S.S.M.'s work is supported by DOE grant DE-FG03-95ER14499.

\newpage
\begin{figure}
\caption{Temporal dependence of the concentrations of $5$ types of binary
molecules inside the reactor flux, replicating without error. 
(a)Segregation: depending on the value of $W_{ii} - D_{i} - \Phi _{0}$, 
the number of molecules grows or dies out exponentially. 
(b)Selection: once the restriction of keeping the population constant inside 
the reactor is applied, only the sequence with the highest productivity 
$W_{ii} - D_{i}$ survives. In (a) $\Phi _{0}=1.0=$ constant. For both 
plots, the other parameters are: $D_{i}=0.5$, $W_{11}=1.9$, $W_{22}=1.8$%
, $W_{33}=1.7$, $W_{44}=1.2$ and $W_{55}=0.5$.}
\label{fidom1}
\end{figure}

\begin{figure}
\caption{Steady-state frequency distribution of mutants $y_{d}$ as a
function of the Hamming distance $d$ from the master sequence,
for four distinct values of the mutation rate: (a)$\mu =0.01$, (b)$\mu =0.04$ 
(c)$\mu =0.07$ and (d)$\mu =0.10$. Other parameters are: $L=30$ and $a=10$.}
\label{fidom2}
\end{figure}

\begin{figure}
\caption{The mortality rate derived from a simulation of the Penna model is 
shown. The exponential behaviour, which appears as a straight line in this 
semi-log plot, shows the agreement with Gompertz law when only genetic 
deaths are taken into account. Crosses: genetic deaths only; Diamonds: deaths 
due to Verhulst are also included. Parameters: $T=3$, $R=8$, $B=1$, $M=1$ 
and initial population around $10^9$ individuals.}
\label{figomp}
\end{figure}

\begin{figure}
\caption{The Figure shows the (normalized) survival rate as a function of 
age for an asexual semelparous population and two iteroparous ones, labeled 
as (a) and (b). In (a), females cease reproduction - undergo menopause - at 
age $13$, and in (b) there is no menopause. In all three $Minage=10$, $B=2$, 
$T=1$ and $M=1$.}
\label{fisalmon}
\end{figure}

\begin{figure}
\caption{The procedure used in the model to simulate each reproductive regime 
is showed through simple examples.}
\label{firegi}
\end{figure}

\begin{figure}
\caption{Same as in Fig. \ref{fisalmon}, now for sexual populations. Sex 
causes the females to have a pos-reproductive life.}
\label{fisurvsex}
\end{figure}

\begin{figure}
\caption{Distribution of female menopause age for two distinct simulations: 
with reproductive risk and maternal care (filled circles) and neither 
reproductive risk nor maternal care (open diamonds). For the first case one 
observes that the age of menopause self-organizes showing a peak at age 20. 
The distribution in the second case is an artifact coming from the 
impossibility of the menopause age to be greater than 32. $Apc$ measures the 
number of time periods of maternal care needed for an infant to survive.}
\label{fimeno}
\end{figure}

\begin{figure}
\caption{Evolution of the total population, after a genetically stable state 
has been reached. At step 50000 the catastrophe kicks in, by setting to $1$ 
a given locus of all the genomes. Notice the extinction of both the haploid 
asexual (AR) and the meiotic parthenogenetic (MP) populations, whereas the 
sexually reproducing one (SR) quickly adjusts itself to the new paradigm.}
\label{ficat}
\end{figure}

\begin{figure}
\caption{The fraction of females that reproduce sexually in the population is 
plotted against the value of the exposure parameter $E$. The correlation 
between the dominant pattern of reproduction and the intensity of the 
infestation, as measured by this last parameter, is clearly seen.}
\label{firedq}
\end{figure}

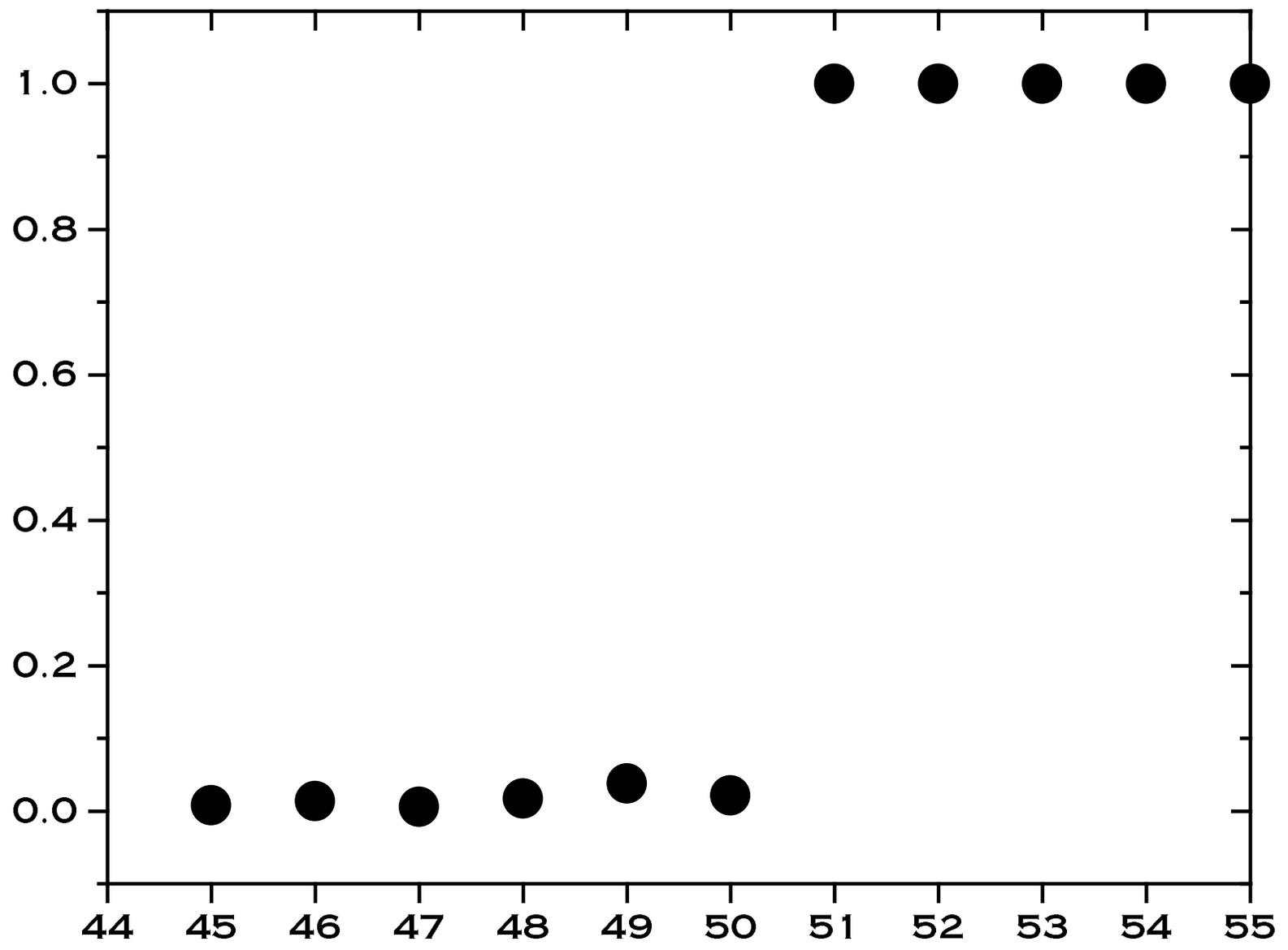
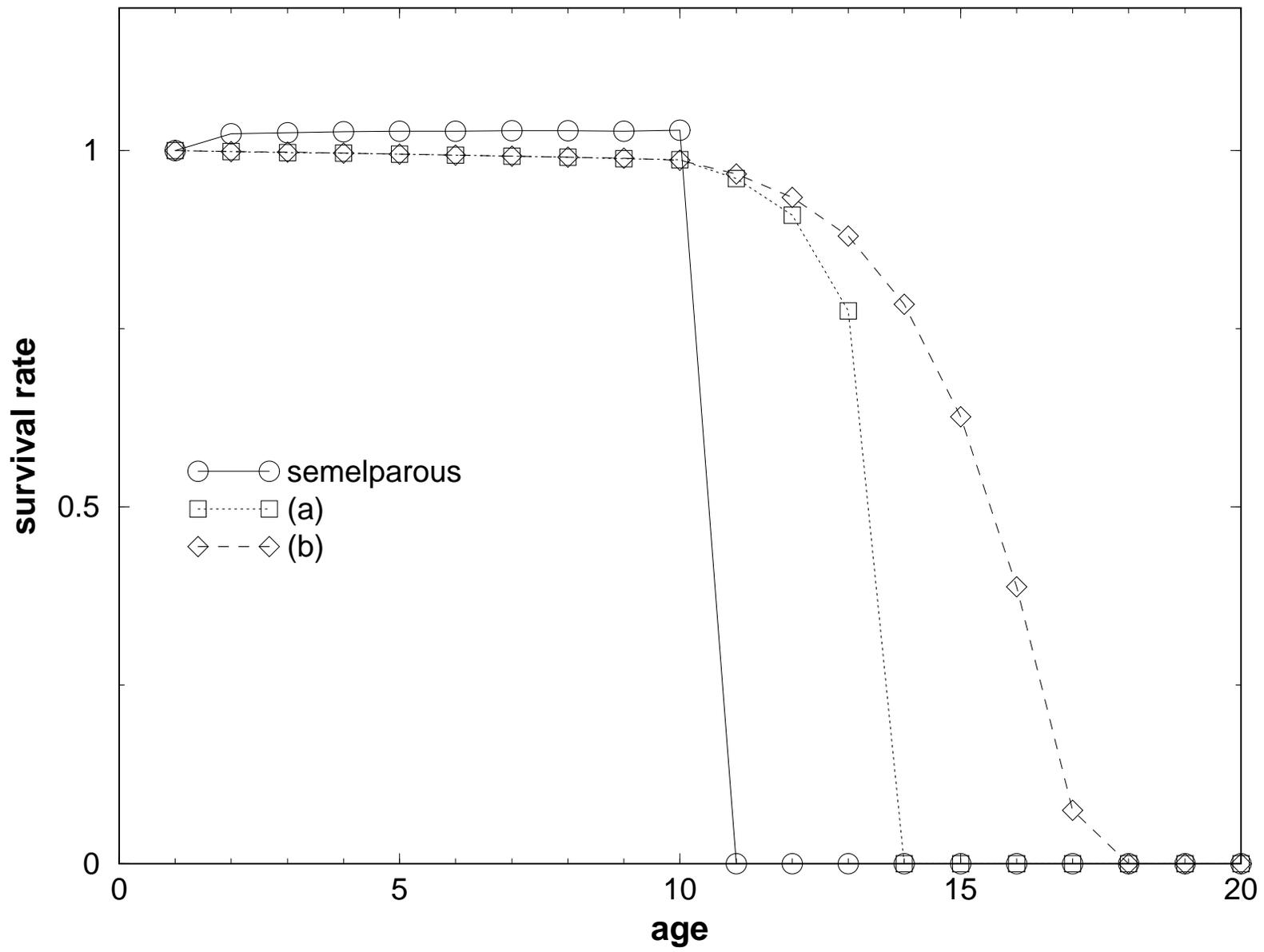
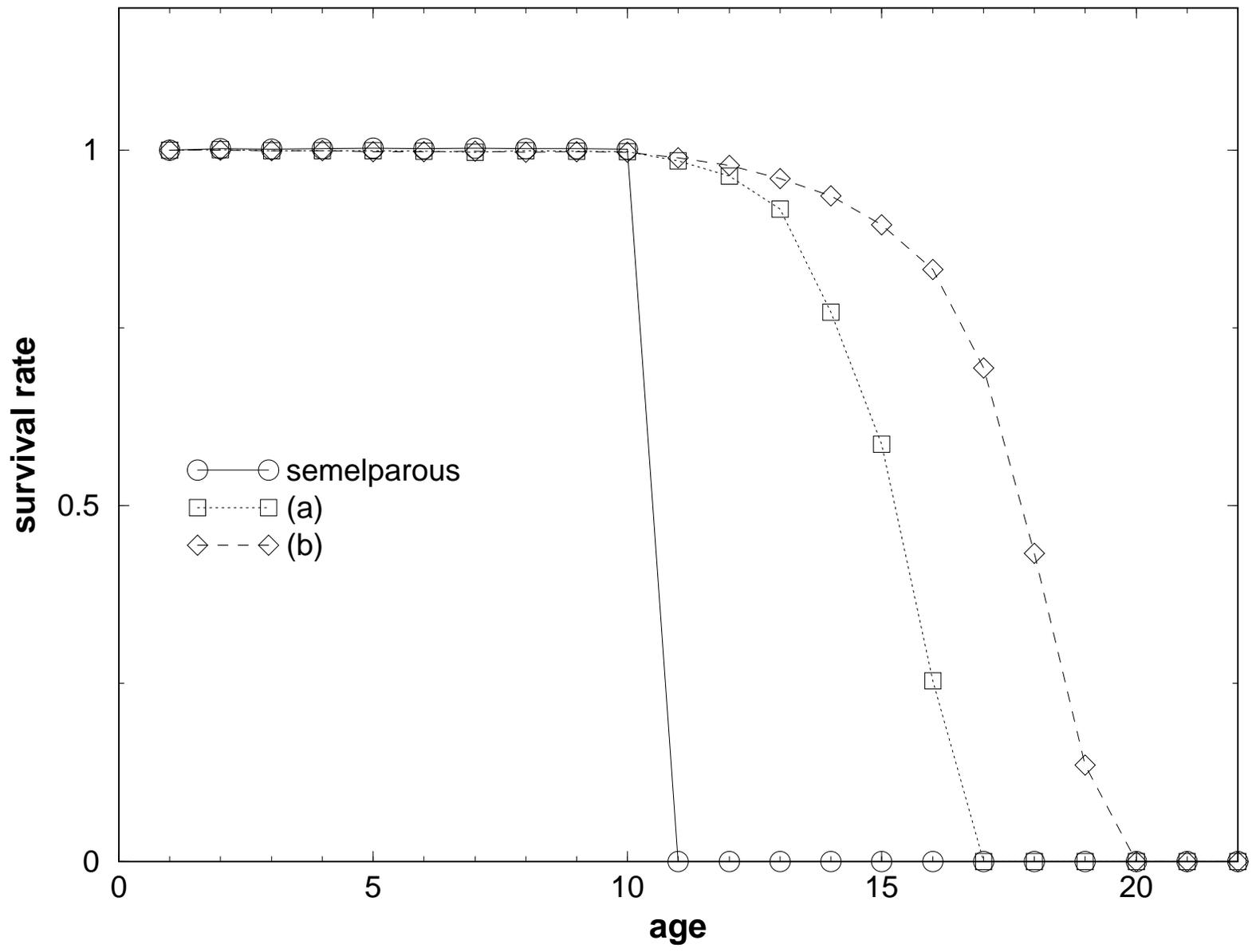
\end{document}